\documentclass[11pt,twoside]{article}

\usepackage{asp2006}
\usepackage{epsf}
\usepackage{psfig}
\usepackage{lscape}
\usepackage{graphicx}

\begin{document}
\title{Asteroseismic age and radius of Kepler stars}
\author{Orlagh L. Creevey}
\affil{Instituto de Astrof\'isica de Canarias, 
C/ V\'ia Lactea s/n, La Laguna 38205, Tenerife, Spain}
\affil{High Altitude Observatory, 30801 Center Green Dr., Boulder, CO, 80301}

\begin{abstract} 

The Kepler mission's  primary goal is the detection
and characterization of Earth-like planets by observing continuously 
a region of 
sky for a nominal period of three-and-a-half years.  
Over 100,000 stars will be monitored, with a small subset of these 
having a cadence of 1 minute, making asteroseismic studies for many 
stars possible. 
The subset of targets will consist of mainly solar-type and 
planet-hosting stars, and these will be observed for a minimum 
period of 1 month and a maximum depending on the scientific yield 
of the individual target.   
Many oscillation frequencies will be detected in these data, 
and these will be used to constrain the star's fundamental parameters.
I investigate
the effect that an increase in a) the length of observation
and b) the signal quality, has on the final determination of 
some stellar 
global parameters, such as the radius and the age.
 

\end{abstract}


\section{Method}
\subsection{Precision of oscillation frequencies}
The Kepler mission \citep{bor97,bor04} will yield high-cadence 
photometric time series for many
hundreds to a few thousands of oscillating solar-type stars.
By using any method to extract periodic signals in these light curves, for 
example an FFT or a least-squares fit to sinusoids, the 
{\it pulsation frequencies} 
or {\it oscillation modes} $\nu$ will be detectable.  
The number of frequencies detected for each star 
will depend primarily on the intrinsic brightness
of the star, its distance and the characteristics of its oscillation modes.
Consequently, the errors associated with each of the detected frequencies also
depends on these characteristics.
For data analysis purposes, however, we can quantify the error on each 
frequency $\epsilon_{\nu}$ as a function of signal-to-noise ratio (SNR) in the
power spectrum, 
the duration of observation $T$ and the natural linewidth of the 
frequency $\Gamma$ (related to the damping rate of the oscillations).
\citet{lib91} gives a formula for the estimation of these errors:
\begin{equation}
\epsilon_{\nu}^2 = f(\beta) \frac{\Gamma}{4\pi T}
\end{equation}
where $\beta$ is the inverse of the SNR and varies usually between 0.1 - 1.0.

\subsection{Determining the constraints on global parameters}
In order to determine the global parameters (and subsequently the
 internal structure)
of a star, one needs to compare a set of {\it model observables} {\bf B} 
to a set of {\it observations} {\bf O}.
We use a stellar structure and evolution code {\it ASTEC} 
\citep{jcd82,jcd07a}, coupled to an adiabatic oscillation code {\it ADIPLS} 
\citep{jcd07b}.
This model takes a set of input parameters {\bf P} and generates a stellar 
model based on {\bf P}.  The expected observables {\bf B} can be 
consequently calculated ($L_{\star}$, magnitudes, mean density ...).
One form of determining the stellar global parameters (these are the {\bf P})
is to incorporate some 
algorithm that minimizes the discrepancies between the model observables and 
the observations, 
and then use
a value (such as $\chi^2$) to quantify the fit to the data:
\begin{equation}
\chi^2 = \sum_{i=1}^M \frac{(O_i - B_i)^2}{\epsilon_i^2},
\end{equation}
for each $i = 1,...,M$ observation and 
$\epsilon_i$ represents the errors on each of the $O_i$.

Once we find a robust fit to the data, we can use the following formula
to calculate the uncertainties in each of the model parameters:
\begin{equation}
\sigma_j^2 = \sum_{k=1}^N \frac{V_{jk}^2}{W_{kk}^2} 
\label{eqn:unc}
\end{equation}
for each $j = 1, ..., N$ model parameter.
Here, $V_{k}$ and $W_{k}$ are
the $k^{\rm th}$ components of the singular value decomposition of the matrix
{\bf D}, whose elements $D_{ij}$ 
are given by $\frac{\partial B_i}{\partial P_j} / \epsilon_i$ 
\citep{pre92,bro94,cre07}.

Here we can see that the uncertainties in the parameters such as mass, radius
and age come from the sensitivity of each of the observables with respect to 
each of the parameters (from models), 
{\it and} the associated error on each of these
measurements.  These quantities can change by  
\begin{itemize}
\item including more measurements, such as {\it classical}
observations ($T_{\rm eff}$, $\log g$, $\log Z/X$) or more oscillation 
frequencies 
\item studying a star at various evolutionary 
stages, because the measurement sensitivities 
to each of the parameters changes as the stellar structure changes

\item varying the expected errors on the measurements $\epsilon$.
\end{itemize}
In the first subsection we discussed the frequency errors and their 
dependencies on the photometric time series.
Here we translate these into uncertainties in the global parameters of 
radius and age as a function of quality of data and duration of observations.




\begin{figure}
\begin{center}
\includegraphics[width=0.95\textwidth]{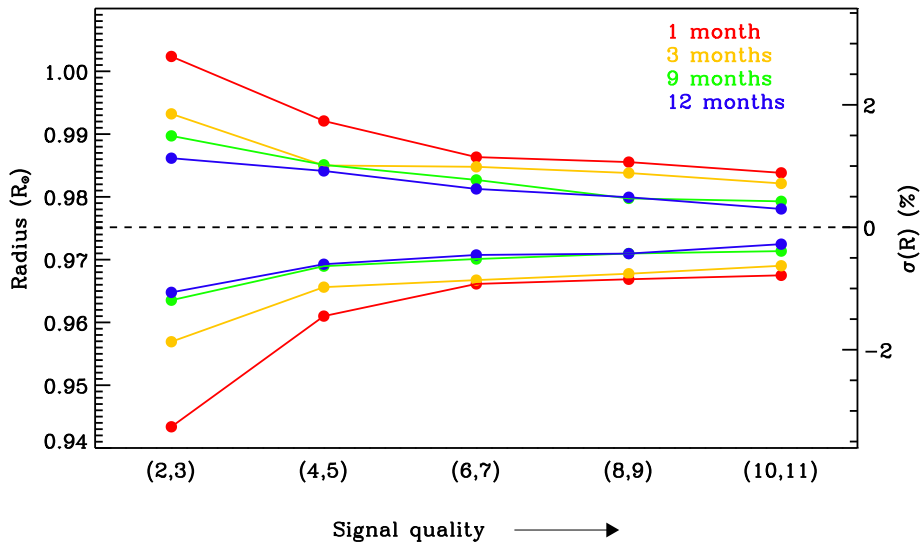}
\end{center}
\caption{Expected uncertainties in the radius of a 1 M$_{\odot}$ star
as a function of {\it signal quality}.  Each of the color curves represents
the results for a specific duration of uninterrupted observations.
The axis on the right shows the relative uncertainties.
\label{fig:poster1}}
\end{figure}

\section{Results}
\begin{figure}
\center{\includegraphics[width=0.95\textwidth]{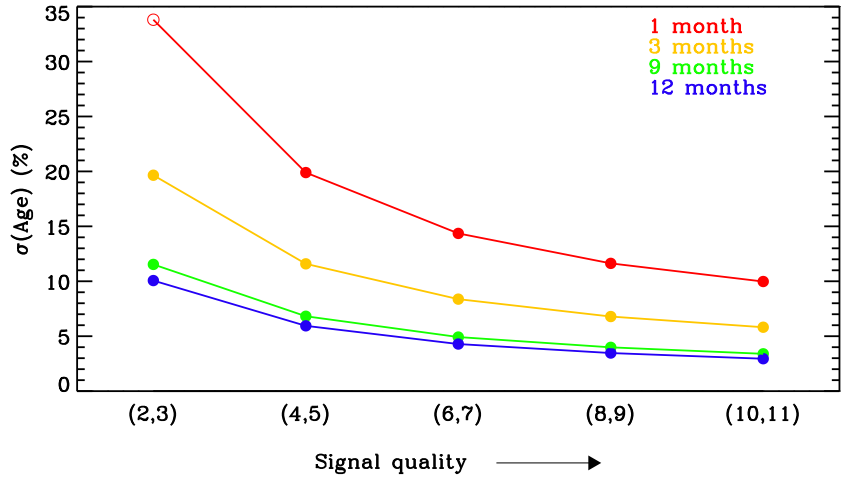}}
\caption{Expected uncertainties in the age of a 1 M$_{\odot}$ star.  See 
Figure~\ref{fig:poster1} for details.  
\label{fig:poster2}}
\end{figure}

We show in Figures \ref{fig:poster1} and \ref{fig:poster2} the expected
uncertainties in the radius and the age, 
respectively,  of a 1 M$_{\odot}$ star as 
a function of {\it signal quality}.
We have quantified {\it signal quality} as (SNR, NF) where 
the 'signal' in SNR refers to the amplitude of the highest 
peak in the spectrum, and the 'noise' refers to the noise level 
around this peak. 
NF is the  number of detected frequencies
for each oscillation mode $l = 0$ and 1 
i.e. NF = 3 means 3 $l$ = 0 modes and 3 $l$ = 1 modes. 

If a star has a SNR of 10, then it is most likely that the 
number of  detected frequencies (NF) in the power spectrum is 
higher than a star whose SNR = 2.
For this reason we may assume that both the SNR and NF increase together.  
These are the numbers on the x-axis in each graph:  
e.g. (SNR, NF) = (2, 3), (4, 5) etc.

Both figures also show the uncertainties in the radius and the age assuming
various {\it lengths of observation}.
The various colour curves represent data of a specific time span 
of high-cadence 100\% duty cycle observations:
red, yellow, green and blue indicate respectively, 1 month, 3 months, 
9 months and 12 months of observational data to extract the oscillation 
frequencies.


\section{Conclusions}
\begin{itemize}
\item The stellar radius can be determined to within 2-3\% for a 1 
M$_{\odot}$ star for all combinations of signal quality and observation
length studied.

\item  Observing a star that has at least 5 detected frequencies for $l$=0 
and $l$=1 modes yields $\sigma(\tau) <$20\%, where $\tau$ is the age, 
and requiring that 
the length of observation is at least 3 months, improves this to better than
12\%.

\item The most significant improvements in $\sigma(R)$ and $\sigma(\tau)$ come
from extending a data set from 1 month to 3 months.
If the data set spans 9 months, then extending it will result in 
very little improvement in these quantities.

\item A star observed for 1 month requires a minumum (SNR, NF) = (6, 7) 
to constrain adequately the radius and the age of a 1 M$_{\odot}$; 
3 months requires a minimum (SNR, NF) = (4, 5); 
9 months requires a minimum (SNR,~NF)~=~(2,~3).

\end{itemize}

\acknowledgements 

OLC wishes to acknowledge Thierry Appourchaux, William J. Chaplin, 
Guenter Houdek, 
Sebastian Jim\'enez-Reyes, Hans Kjeldsen, Travis Metcalfe and David Salabert for discussions on various aspects of this work.  OLC also wishes to acknowledge the ISSI for financial support for the ``AsteroFLAG'' meeting, where this idea was elaborated upon, and travel support was provided by the AAS International Travel Grant. This research was in part supported by the European Helio- and Asteroseismology Network (HELAS), a major international collaboration funded by the European Commission's Sixth Framework Programme.


\end{document}